\documentclass[aps, prx,reprint, superscriptaddress]{revtex4-1}
\usepackage{graphicx}
\usepackage{amsmath, color, ulem}
\usepackage{dsfont, bm,  units}

\usepackage{soulutf8}%erlaubt das farbliche markieren des Textes
\usepackage{upgreek}

\def  \Bxc       {{B_{\rm xc}}}      % exchange correlation potential
\def  \Gup       {{G^{\uparrow}}}    % up-spin Green function
\def  \Gdn       {{G^{\downarrow}}}  % down-spin Green function
\def  \br        {{\bm r}}           % bold r (position vector)

\begin{document}

\title{High-frequency magnon excitation due to femtosecond spin-transfer torques}
\author{Ulrike Ritzmann}
\affiliation{Department of Physics and Astronomy, Uppsala University, Box 516, 75120 Uppsala, Sweden}
\affiliation{Dahlem Center of Complex Quantum Systems and Department of Physics, Freie Universit\"at Berlin, Arnimallee 14, 14195 Berlin, Germany}
\author{Pavel Bal\'a\v{z}}
\affiliation{Charles University, Faculty of Mathematics and Physics, Department of Condensed Matter Physics, Ke Karlovu 5, CZ 121 16 Prague, Czech Republic}
\affiliation{IT4 Innovations Center, VSB Technical University of Ostrava, 17. listopadu 15, CZ 708 33, Ostrava-Poruba, Czech Republic}
\author{Pablo Maldonado}
\affiliation{Department of Physics and Astronomy, Uppsala University, Box 516, 75120 Uppsala, Sweden}
\author{Karel Carva}
\affiliation{Charles University, Faculty of Mathematics and Physics, Department of Condensed Matter Physics, Ke Karlovu 5, CZ 121 16 Prague, Czech Republic}
\author{Peter M. Oppeneer}
\affiliation{Department of Physics and Astronomy, Uppsala University, Box 516, 75120 Uppsala, Sweden}
\date{\today}

\begin{abstract}
Femtosecond laser pulses can induce ultrafast demagnetization as well as generate bursts of hot electron spin currents. In trilayer spin valves consisting of two metallic ferromagnetic layers separated by a nonmagnetic one, hot electron spin currents excited by an ultrashort laser pulse propagate from the first ferromagnetic layer through the spacer reaching the second magnetic layer. When the magnetizations of the two magnetic layers are noncollinear, this spin current exerts a torque on magnetic moments in the second ferromagnet. 
Since this torque is acting only within the sub-ps timescale, it excites coherent high-frequency magnons as recently demonstrated in experiments.
Here, we calculate the temporal shape of the hot electron spin currents using the superdiffusive transport model and simulate the response of the magnetic system to the resulting ultrashort spin-transfer torque pulse by means of atomistic spin-dynamics simulations. 
Our results confirm that the acting spin-current pulse is short enough to excite magnons with frequencies beyond $1\, {\rm THz}$, a frequency range out of reach for current induced spin-transfer torques. 
We demonstrate the formation of thickness dependent standing spin waves during the first picoseconds after laser excitation. In addition, we vary the penetration depth of the spin-transfer torque to reveal its influence on the excited magnons. 
Our simulations clearly show a suppression effect of magnons with short wavelengths already for penetration depths in the range of $1\, {\rm nm}$ confirming experimental findings reporting penetration depths below $2\, {\rm nm}$. 
%Here (for the first time) we simulate the response of magnetic system to such sub-ps spin-transfer torque pulse by means of atomistic spin dynamics. Our calculations confirm that these times are short enough to excite coherent magnons with frequencies reaching 1 THz, a range out of reach for current induced spin-transfer torque. Recently,  such generation of high frequency magnons has been demonstrated experimentally. It was possible to draw  conclusions about the spin penetration depth from the observation of higher standing wave modes in the spectrum.  Here we directly relate the distribution of higher frequency modes and the spin penetration depth. 
%A critical component is the temporal shape of the spin current pulse, which we obtain from the superdiffusive spin transport model. Our method thus allows to examine the effect of the trilayer composition on the generated magnon spectrum.

\end{abstract}

\pacs{} 
%\keywords{}

\maketitle

\section{Introduction}
\label{Sec:Introduction}
The first experimental observation of ultrafast demagnetization due to femtosecond laser excitation in nickel were reported more than twenty years ago~\cite{Beaurepaire1996}. 
Since then a variety of research activities has focused on studying the magnetization dynamics induced by intense, ultrashort laser pulses~\cite{Walowski2016,Carva2017, Malinowski2018}. 
For the purpose of possible technological applications, especially, all-optical magnetization switching has become a topic of current research~\cite{Stanciu2007, Radu2011, Lambert2014, Lalieu2017a, Wilson2017}.
However,  the relevant microscopic scattering processes and their interplay leading to ultrafast demagnetization are still under debate \cite{Koopmans2005, Carpene2008, Koopmans2010,  Battiato2010, Fahnle2011, Carva2011, Carva2013, Krieger2015}.
Laser-induced ultrafast demagnetization can also lead to spin-polarized currents of hot electrons   ~\cite{Battiato2010,Malinowski2008,Melnikov2011, Rudolf2012, Alekhin2017}.
Moreover, experiments have suggested that a single pulse of hot electron spin currents without any assistance of laser heating induces ultrafast demagnetization of an adjacent magnetic layer~\cite{Eschenlohr2013, Vodungbo2016, Bergeard2016, Xu2017}.
Importantly, it has been demonstrated \cite{Schellekens2014, Choi2014} that femtosecond spin currents carried by hot electrons can exert spin-transfer torque (STT)~\cite{Slonczewski1996, Berger1996, Slonczewski2002}
on a ferromagnet leading to the rotation of magnetization or the excitation of high frequency magnons~\cite{ Razdolski2017} in a confined ferromagnetic structure. 

Spin and energy resolved transport of laser-excited hot electrons can be described by the model of superdiffusive transport introduced by Battiato {\it et al.}~\cite{Battiato2010, Battiato2012}.
The model is based on semiclassical equations of motion for electrons having nonthermal energies above the Fermi level and moving in two spin channels. %with different energies located in a certain range.
%UR: Which range? Maybe be more specific?
A strong asymmetry of the electron velocities and life times for the different spin channels in ferromagnetic metals leads to spin-polarized currents, following electron excitation, which can propagate across the nonmagnetic layer within femtoseconds and enter another magnetic layer. Alternatively, spin-dependent transport of hot electrons can also be described by a model based on the Boltzmann equation ~\cite{Nenno2016, Nenno2018}, which leads to similar results. 
%If the magnetizations of both magnetic layers are collinear, magnetization of the second magnetic layer remains uninfluenced. 
% Comment Ulrike: Are you sure that statement generally holds? It is only a weak effect, but still something happens?
If the magnetizations of both magnetic layers are noncollinear, the transverse part of the spin current (with respect to the local magnetization direction) will be absorbed by the ferromagnet and transformed into an STT inducing magnetization dynamics~\cite{Stiles2002, Barnas2005, Balaz2018}. 
These femtosecond STT due to hot electron spin currents can trigger the excitation of high-frequency magnons, which form standing spin waves in ultrathin magnetic layers~\cite{Razdolski2017, Ulrichs2018}.

Here, we present a theoretical study combining the superdiffusive spin-transport and atomistic spin-dynamics simulations~\cite{Nowak2007, Skubic2008, Evans2014} to describe the ultrafast magnon excitation in a metallic spin valve.
We start by calculating the laser-excited spin current in the first ferromagnetic layer via the superdiffusive spin transport theory~\cite{Battiato2014} extended to describe the spin current propagating through the spacer and considering perpendicular alignment of the two ferromagnetic layers~\cite{Balaz2018}. 
The excited spin current exerts STT on the second ferromagnetic layer leading to magnetization dynamics within this layer.  We describe this magnetization dynamics with atomistic spin-dynamics simulations. Particularly, we discuss the thickness dependent magnon frequency spectra considering atomistic spin dynamics with exchange interactions beyond the nearest neighbor approach with exchange parameters determined by \textit{ab initio} methods. 
Furthermore, it has been shown experimentally that the components of the spin current transverse to the layer's magnetization are not absorbed directly at the interface between the nonmagnetic and magnetic layers, rather, it penetrates into the magnetic layer up to a distance of few nanometers~\cite{Razdolski2017, Lalieu2017}.
Therefore, we consider the STT that not only acts on the first atomic layer of the magnetic material, but that affects also subsequent magnetic moments taking into account gradual decrease of STT. Consequently, we study STT effect on the magnon spectra dependent on the considered penetration depth of the transverse spin current. Lastly, we analyze the time evolution of the frequency spectra. Our results reveal further insights in how to tailor the trilayer composition so that the desired magnonic contribution is enhanced.

This paper is organized as follows. In Sec.~\ref{Sec:Methods} we introduce the methods we have used. 
We outline both the basic approach to spin-torque calculation as well as atomistic spin-dynamics simulations.
In Section \ref{Sec:Results} we present a description and discussion of our results. 
Finally, we conclude in Sec.~\ref{Sec:Conclusions}.

\section{Methods}
\label{Sec:Methods}
%%%
%In our calculations, we assume a spin valve type magnetic trilayer of a structure FM1($d_1$)$|$NM($d$)$|$FM2($d_2$), where FM1 and FM2 are ferromagnetic layers and NM is a nonmagnetic spacer. The setup is illustrated in Figure \ref{Fig:device}.
%The numbers in the brackets corresponds to widths of the layers.
%For both ferromagnetic layers we consider the same material, Fe, while the nonmagnetic spaces is assumed to be made of Cu. 
%The trilayer system is excited by an ultrashort laser pulse. In a first step, we describe the used model to calculate the hot electron spin current and the resulting STT that is acting on the second magnetic system. In a second part, we introduce the concepts of the performed atomistic spin dynamics simulations describing the excited magnetization dynamics in the second ferromagnet.

\subsection{Femtosecond spin-transfer torque terms due to laser excitation in FM1$|$NM$|$FM2 trilayers}
\begin{figure}[t]
    \centering
    \includegraphics[width=.98\columnwidth]{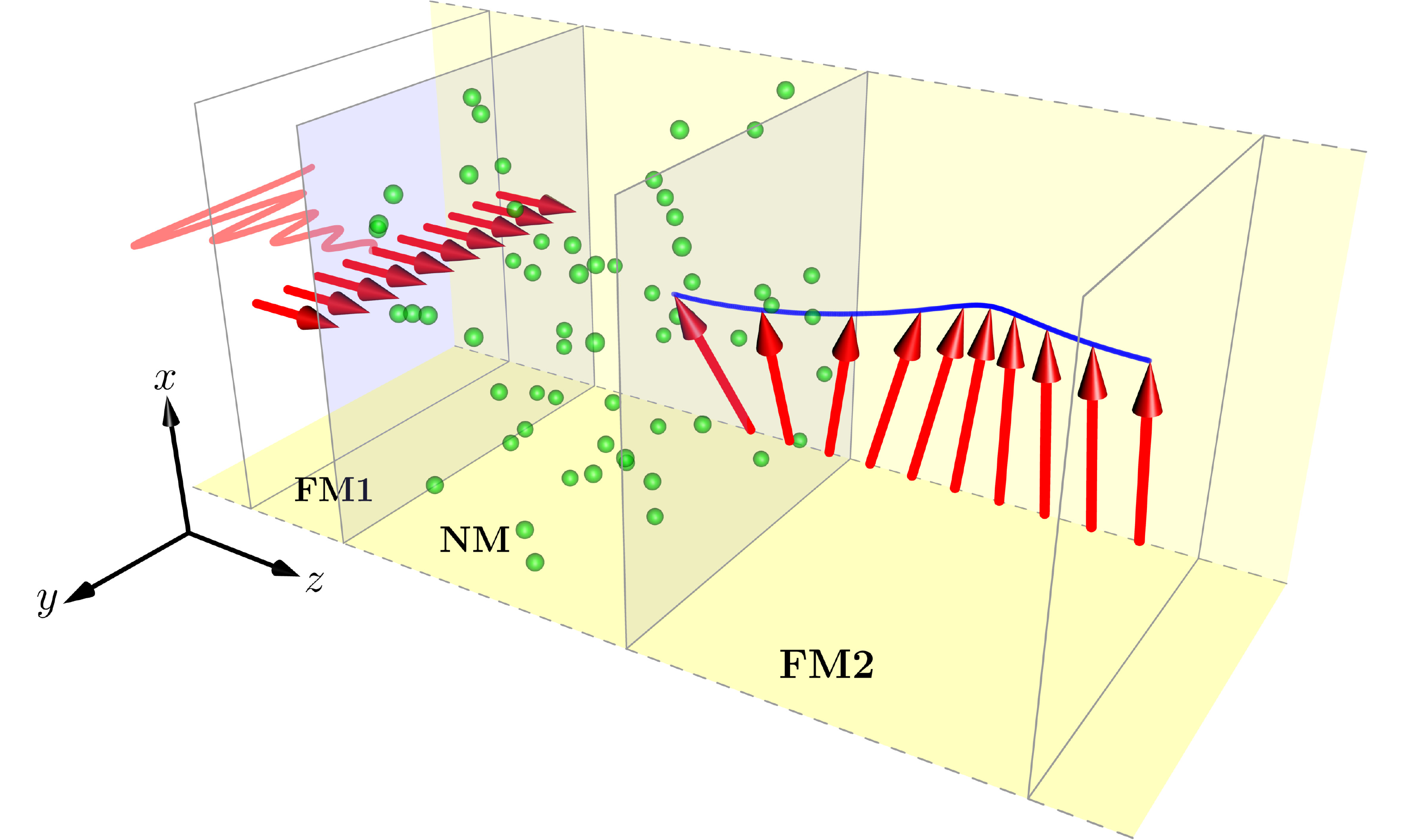}
    \caption{\label{Fig:device} (Color online) Studied trilayer structure FM1$|$NM$|$FM2 made of two ferromagnetic layers (FM1 and FM2) separated by a nonmagnetic one (NM). Magnetic moments are presented by the red arrows. Layer FM1, which is assumed to have perpendicular magnetic anisotropy, 
    is excited by a femtosecond laser pulse from the back of the device. A spin current of hot electrons (green balls) generated in FM1 is transmitted through the NM layer into FM2, where it exerts a spin-transfer torque on the magnetic moments causing spin waves excitations.}
\end{figure}

In our calculations, we assume a spin-valve type magnetic trilayer of a structure FM1($d_1$)$|$NM($d$)$|$FM2($d_2$), where FM1 and FM2 are metallic ferromagnetic layers and NM is a metallic nonmagnetic spacer. The setup is illustrated in Figure \ref{Fig:device}.
The numbers in the brackets corresponds to the widths of the layers. In our calculations we have used $d_1=16$\,nm and $d=4$\,nm and we have varied the thickness $d_2$.
For both ferromagnetic layers we consider the same material, Fe, while the nonmagnetic spacer is assumed to be made of Cu. 
These trilayer system is excited by an ultrashort laser pulse acting on the left side of FM1 and we describe the resulting spin and energy resolved transport of hot electrons using the superdiffusive-transport model as introduced by Battiato {\it et al.}~\cite{Battiato2010, Battiato2012}.

The main input parameters of the model are energy and spin dependent electron velocities and life times, which are obtained from {\it ab initio} calculations~\cite{Zhukov2005, Zhukov2006}.
%UR: Why is this important? Either discuss with details or remove it. I suggest removing..
%Importantly, the model demonstrates that the type of the electronic transports changes from ballistic to diffusive. This changes happens on a timescale of few picoseconds during which the electronic transport remains superdiffusive~\cite{Battiato2010:PRL}.
%UR: Rewritten! We have focused on a situation when a femtosecond laser pulse is applied from the left hand side of the trilayer. 
We consider laser excitation of only the first ferromagnet.
Consequently, the electrons from the $d$ band are excited into the $sp$-band above the Fermi level.
In our calculations, we have assumed 12 uniformly distributed energy levels above the Fermi level with energy spacing $\Delta\epsilon = 0.125\, {\rm eV}$. Thus electrons up to $1.5\,{\rm eV}$ above the Fermi level are excited by the laser pulse. 
We assume that the laser pulse populates the same electron density on each energy level. 
Moreover, the same amount of electrons is populated in both spin channels.

The initial time dependence of the hot electron distribution is given by the temporal shape of the laser pulse.
We consider a Gaussian-shaped laser pulse, which dictates the time-dependent number of electrons on each energy level, $\epsilon$, in spin channel, $\sigma \in \{ \uparrow, \downarrow\}$, being evolved as
\begin{equation}
    N_\sigma(t,z,\epsilon) = \bar{N}_\sigma(\epsilon,z)\; \frac{1}{\Delta \sqrt{2\pi}}
    \exp \left\{ \frac{(t - t_0)^2}{2\, \Delta^2} \right\}\,,
\label{Eq:N_time}
\end{equation}
where $\bar{N}_\sigma(\epsilon,z)$ is the average number of excited electrons at energy level $\epsilon$ with spin $\sigma$ at position $z$. Moreover, $t_0$ stands for the time-zero position of the pulse peak while $\Delta$ is the pulse width.
Here we have assumed $\Delta = 40\, {\rm fs}$.
In order to assume a finite penetration width of the laser, $\ell$, we assume that the average number of excited electrons exponentially decreases with the distance from the left interface of FM1 layer located at $z = 0$
\begin{equation}
%Ulrike: Where do we define z=0? I always consider z=0 at Interface NM|FM2? How can we define these things consequently?
    \bar{N}_\sigma(\epsilon,z) = \bar{N}_0\, \exp\left( -z / \ell \right)\,,
\label{Eq:N_z}
\end{equation}
where $\bar{N}_0 = \bar{N}(0, \epsilon)$ is the same for all energy levels, $\epsilon = \epsilon_i$. 
In our calculations we have used $\bar{N}_0 = 0.1$ electrons per level, corresponding to a laser fluence $F = 27.5\, {\rm mJ}\cdot {\rm cm}^{-2}$ for Fe. The laser penetration depth is assumed to be $\ell = 15\, \mathrm{nm}$ in all layers.
Moreover, the model assumes both gradual relaxation of high energy electrons toward lower energy levels as well as generation of higher order electrons due to elastic scattering of itinerant electrons on atoms.

\begin{figure}[t]
    \centering
    \includegraphics[width=0.49\textwidth]{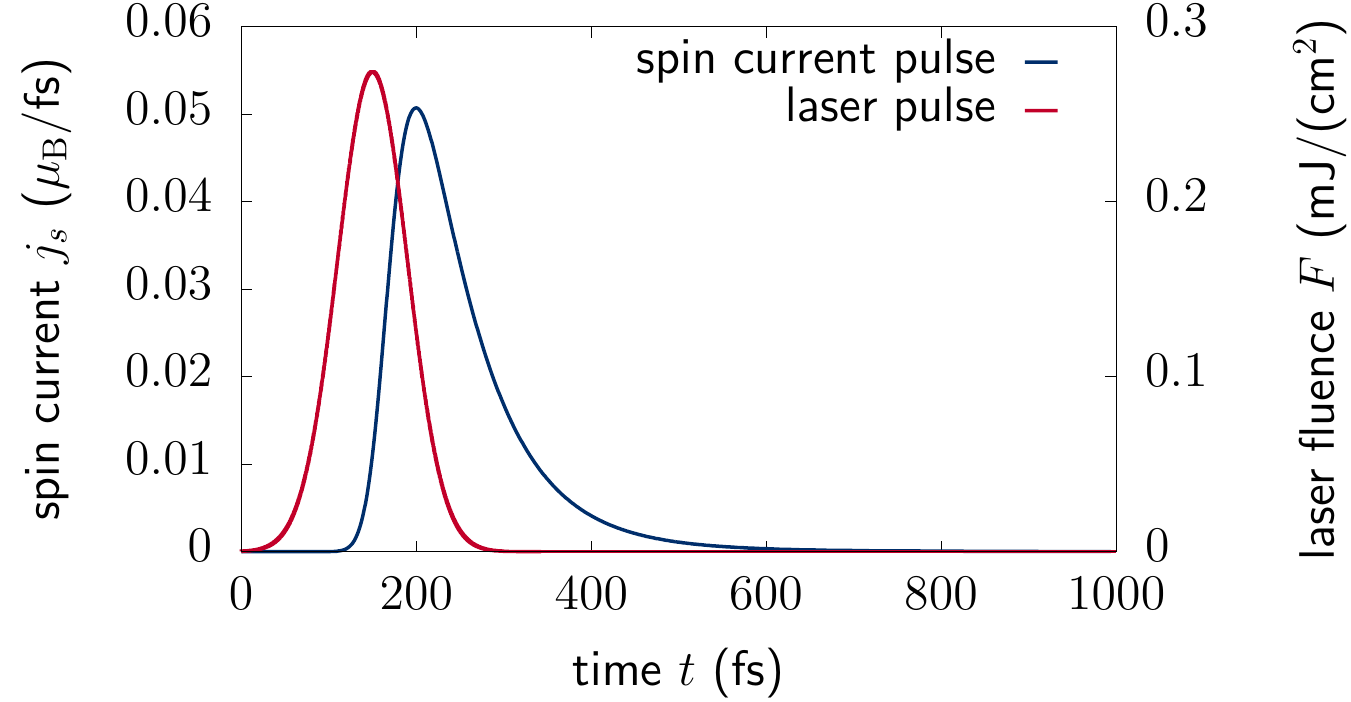}
    \caption{\label{fig2} (Color online) Laser fluence of the laser pulse and calculated superdiffusive hot-electron spin current $j_s$ per atom entering the second FM as a function of time $t$. Note that the maximum of the laser pulse is at $t_0=150$\,fs.}
\end{figure}

Importantly, we consider the FM2 magnetization being perpendicular to the one of FM1. Furthermore, we suppose that the spin current generated and polarized by FM1 layer is entirely absorbed by the FM2 layer and thus completely transformed into spin-transfer torque. This gives rise to a damping-like spin-torque term exerted on magnetic moments. For more details on the spin-torque calculation methodology, see Ref.~\onlinecite{Balaz2018}. The resulting spin current entering the second ferromagnet has a duration of about $500\, {\rm fs}$ and its calculated time-dependent profile is shown in Figure~\ref{fig2}.
%Ulrike: Are the units okay? Is j_s=\tau_{IF}? Please check!

\subsection{Atomistic spin-dynamics simulations}

For our spin-dynamics simulations we consider a thin iron layer with a bcc lattice. 
To describe the resulting dynamics we solve the Landau-Lifshitz-Gilbert (LLG) equation numerically \cite{Nowak2007}
\begin{align}
 \frac{\partial \mathbf{m}_i}{\partial  t}=&-\frac{\gamma}{\mu_\mathrm{s}} \mathbf{m}_i(t) \times \mathbf{H}_i(t)+\alpha\, \mathbf{m}_i(t) \times \frac{ \partial\mathbf{m}_i (t)}{\partial t}\nonumber\\  &+\; \frac{\tau_{\mathrm{IF}}}{\mu_\mathrm{s}}(t,z)\, \mathbf{m}_i(t)\times \mathbf{m}_i(t) \times \mathbf{\hat{z}}\mathrm{,}
\end{align}
for the (normalized) magnetic moments $\mathbf{m}_i$ of each atom in the iron layer at lattice position $i$. The equation consists of a precessional term of the normalized magnetic moment $\mathbf{m}_i=\mathbf{M}_i/\mu_\mathrm{s}$ around its effective field $\mathbf{H}_i$ and a phenomenlogical relaxation term with damping constant  $\alpha$~\cite{Gilbert2004}. 
Moreover, we consider an additional dissipative term due to the femtosecond STT that acts like an anti-damping torque term.  
$\gamma=1.76\cdot10^{11}$\,T/s denotes the gyromagnetic ratio, $\mu_\mathrm{s}=2.2\, \mu_\mathrm{B}$ is the absolute value of magnetic moment of each atom and $\mu_\mathrm{B}=9.27\cdot10^{-24}$\,J/T is the Bohr magneton. 
The effective field $\mathbf{H}_i(t)$ is given by the derivative of the Hamiltonian with respect to the magnetic moment $\mathbf{m}_i$. We consider exchange interaction beyond the nearest neighbor approach, and an anisotropy including crystalline anisotropy, as well as a shape anisotropy given by the demagnetization field, and use the following Hamiltonian:
\begin{align}
	%H = &-\sum_{ij} J_{ij} \mathbf{m}_i \cdot \mathbf{m}_j -\sum_{i} \mathbf{B} \cdot \mathbf{m}_i \\ \nonumber & - \sum_{i} \left(d_x \left(m_i^x\right)^2+ d_z \left(m_i^z\right)^2\right)
	\mathcal{H} =-\sum_{i<j}J_{ij}\, \mathbf{m}_i \cdot \mathbf{m}_j - \sum_{i}\big(k_x \left(m_i^x\right)^2+ k_z \left(m_i^z\right)^2\big)
\end{align}
The exchange interaction between the magnetic moments at lattice nodes $i$ and $j$ in iron can alternate in sign depending on the distance of the two magnetic moment to each other leading to frustration effects.
The values of the exchange interactions and details of their calculation can be find 
in Appendix~\ref{App:exch_int}. The exchange interaction in metals like iron is long range. In our numerical simulations, we include exchange interactions up to the 6th neighboring shell, which corresponds to the distance of $2a$, where $a$ is the lattice constant. Note, that exchange interactions at higher neighbors might be still relevant and affect the dispersion relation and the effective spin wave stiffness. However, this will only cause a minor shift of the frequencies of the standing waves, but it will not further have an effect on the results.

Due to the small thickness of the second ferromagnet, the demagnetization field due to the dipolar interaction of the magnetic moments causes a shape anisotropy which aligns the magnetic moment perpendicular to the $z$-direction and therefore, perpendicular to
the first magnetic layer. We consider a uniaxial anisotropy with a hard axis in $z$ direction and an anisotropy constant of $k_z=-0.267$\,meV. Furthermore, we consider a magnetic anisotropy aligning the magnetic moments in $x$ direction with an anisotropy constant of $k_x=0.00697$\,meV. Both values are taken from experiments by Razdolski \textit{et  al.} \cite{Razdolski2017}. We initialize the magnetic moments parallel to $x$-direction, which is taken as the magnetic ground state of the system.

The magnetic system is excited by femtosecond STT, which has been determined as described in the previous section. In the following, we use the same STT in all performed simulations, but we couple  the STT in two different ways to the magnetic system.  In our first calculations, we consider that the spin current $j_\mathrm{s}$ per atom is completely absorbed by the first magnetic plane,
\begin{align}
    \tau_\mathrm{IF}(t,z) = %\tau_\mathrm{IF}(t)\, \delta(z)=
    j_\mathrm{s}\cdot\delta(z),
\end{align}
where $z=0$ corresponds to the interface with the normal metal.
In further calculations, we assume spin-current absorption within a characteristic penetration depth $\lambda_\mathrm{STT}$~\cite{Zwierzycki2005,Balaz2013,Whang2019} and therefore, the STT acts on more than one atomic layer and its spatial dependence is given by:
\begin{align}
\label{spatial_STT}
    \tau_\mathrm{IF}(t,z)=%\frac{\tau_\mathrm{IF}(t)}{\sum_z{\exp\left(-\frac{z}{\lambda}\right)}}\, \exp{\left(-\frac{z}{\lambda}\right)}
     \frac{j_\mathrm{s}}{\sum_z{\exp\left(-\frac{z}{\lambda_\mathrm{STT}}\right)}}\, \exp{\left(-\frac{z}{\lambda_\mathrm{STT}}\right)}\,.
\end{align}
We perform atomistic spin-dynamics simulations by numerically integration of the LLG equation using the Heun method \cite{Nowak2007} with a time step of $0.1\,\mathrm{fs}$.
We study bcc-Fe layers with a lattice constant $a=0.287\, \mathrm{nm}$ and different thicknesses $d_2$ ranging from $4.3\, \mathrm{nm}$ to $10\, \mathrm{nm}$. In addition, we consider a cross section of $4.3\,\mathrm{nm} \times 4.3\,\mathrm{nm}$ and apply periodic boundary conditions in $x$ and $y$ directions. These boundary conditions are relevant to avoid effects in $x$ and $y$ directions on the spin-wave dispersion relation.

\section{Results}
\label{Sec:Results}

To start with, we consider a femtosecond STT which is absorbed completely at the interface of the second ferromagnetic layer and we use a thickness of $d_2=25\, a = 7.2\, \mathrm{nm}$ of the second ferromagnet. %where $a$ is Fe lattice constant.
Using the atomistic spin-dynamics simulations we study the resulting magnetization dynamics during the first $100\, \mathrm{ps}$.
The ultrafast STT excites magnetization dynamics leading to the creation of high-frequency magnons which propagate through the second magnetic layer and can be reflected multiple times before decaying.

\begin{figure}[t]
    \centering
    \includegraphics[width=0.49\textwidth]{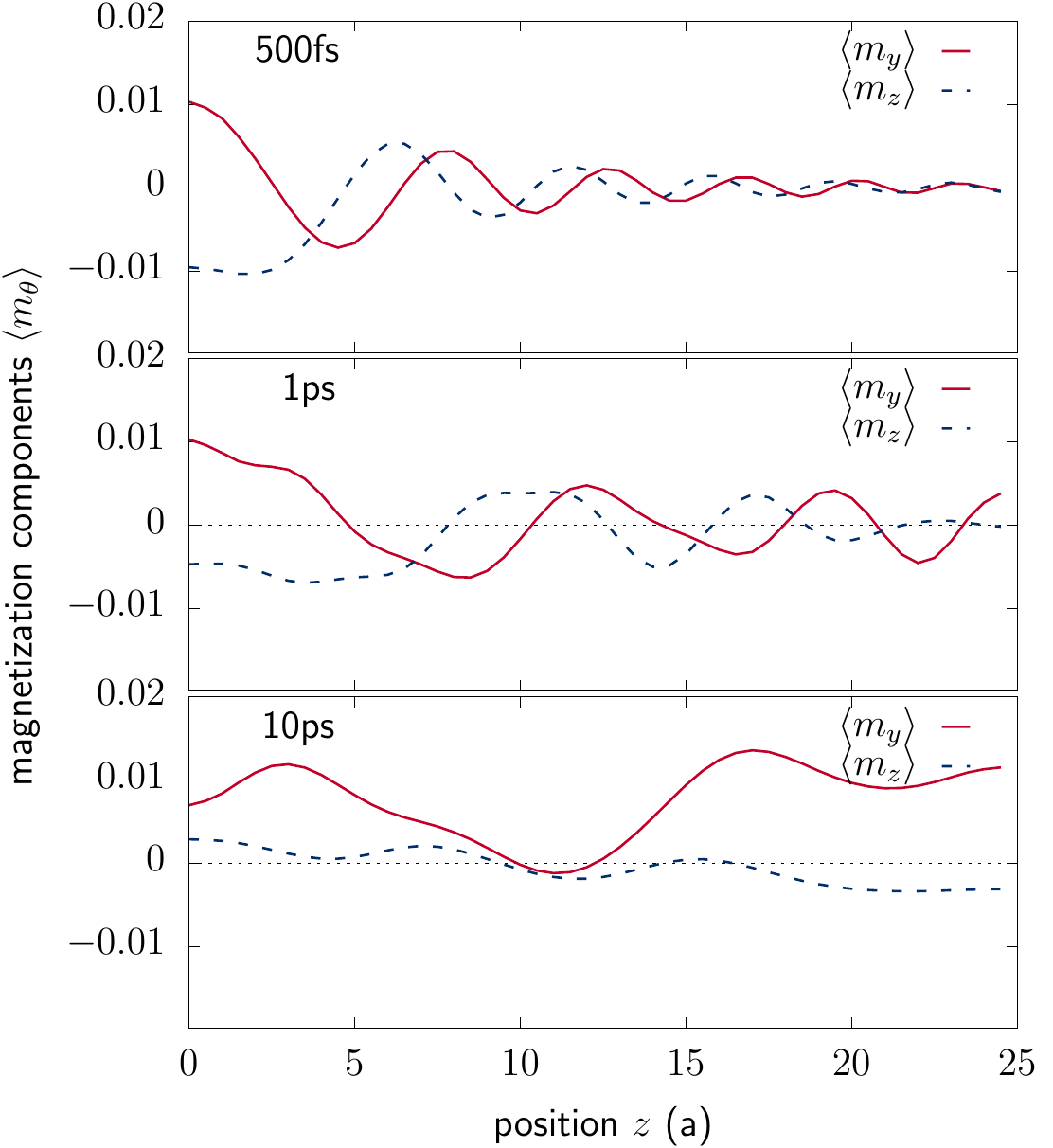}
    \caption{\label{fig3} Calculated spatial profile of the transverse magnetization components at different times.}
\end{figure}

In Figure~\ref{fig3} we show the spatial profile of the magnetization components $\langle m_y \rangle$ and $\langle m_z\rangle$ which are transverse to the initial magnetization direction for different times. $\langle\bullet\rangle$ denotes the average over the plane perpendicular to the propagation direction of the spin waves. At $500\, \mathrm{fs}$ and one can see that the signal is characterized by magnons with short wavelengths and at $1\, \mathrm{ps}$ and $10\, \mathrm{ps}$, one can clearly see that magnons with larger wavelength become more relevant. 
This shows that a broad spectrum of frequencies is excited within the first picoseconds, but mainly magnons with longer wavelengths remain at times beyond $10\, \mathrm{ps}$.
%This shows that a broad spectra of frequencies is excited by ultrafast STT, but mainly magnons with longer wavelengths remain at a period beyond of $10\, \mathrm{ps}$. 
%A reason for that behavior is the frequency dependent lifetime of the magnons, which decreases for higher frequencies. 

\subsection{Formation of standing waves dependent on the thickness}
\begin{figure}[t]
    \centering
    \includegraphics[width=0.49\textwidth]{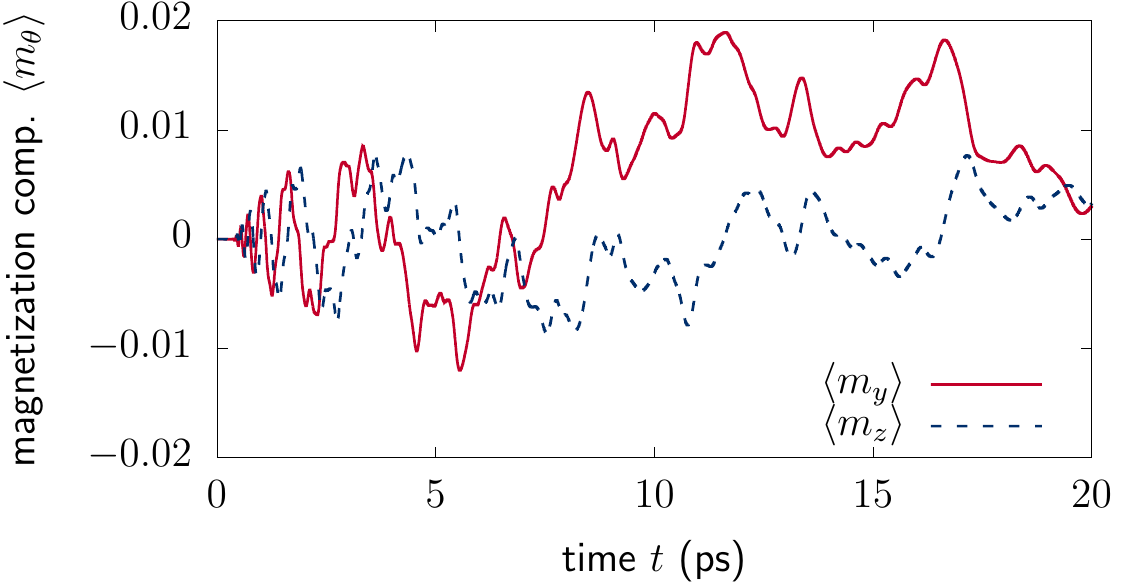}
    \includegraphics[width=0.49\textwidth]{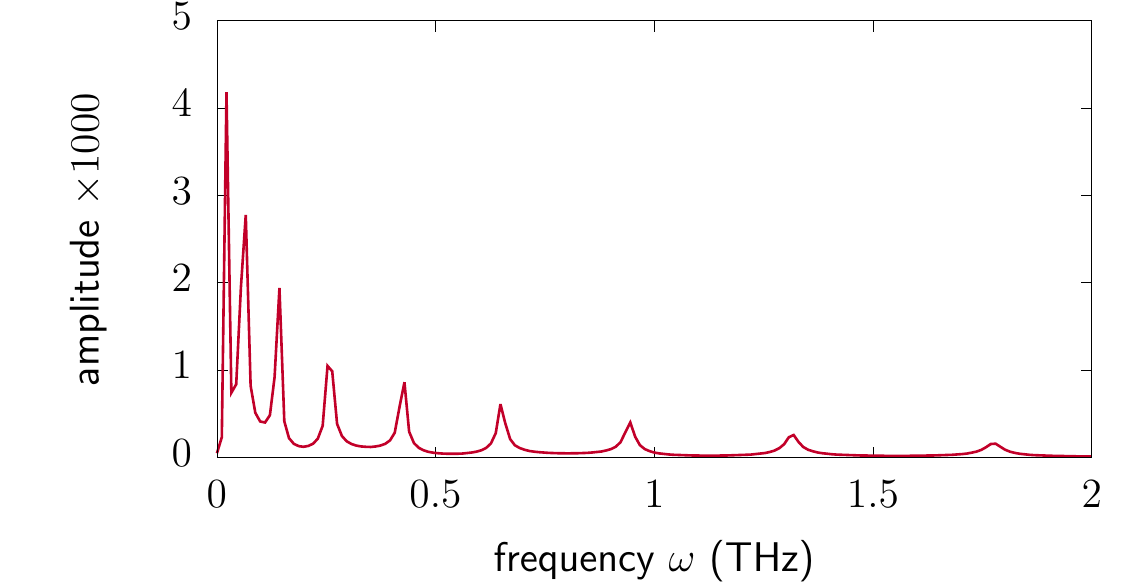}
    \caption{\label{fig4} Frequency spectra of the excited magnetization dynamics. Top: Calculated averaged magnetization components $\langle m_x\rangle$ and $\langle m_y\rangle $ of the last layer in the system. Bottom: Amplitude of the magnons as a function of frequencies obtained by Fourier transformation of the magnetization data.}
\end{figure}

We find furthermore that standing waves are formed due to the small thickness of the ferromagnet. 
Only magnons with wavelengths fitting into the system dimensions remain after a few picoseconds (see below).
For a further analysis, we study the time-evolution of the averaged magnetization of the last layer, $\langle\mathbf{m}(z=d_2)\rangle$ and perform a Fourier transformation in the time domain to determine the appearing frequencies. 
The magnetization components as a function of time are shown in the upper panel of Figure~\ref{fig4}. 
The magnetization components oscillate very fast in the first picoseconds and then slower due to lower frequencies involved afterwards. 
The system remains excited over more than 100\,ps. 
Note that the $y$-component of the magnetization oscillates stronger than the $z$-component. 
This is a consequence of the hard axis of the anisotropy in $z$-direction, which suppresses larger amplitudes in that magnetization direction.

In the lower panel of Figure \ref{fig4}, the corresponding magnon amplitudes are shown as a function of frequencies. 
To obtain the amplitudes we perform a Fourier transformation of $m^+ = \langle m_x\rangle+i \langle m_y\rangle$ of the averaged magnetization in the last layer. 
%Note that due to the strong anisotropy providing a hard axis in the $z$-direction, the magnetization components in $x$- and $y$- direction are not necessarily equal. 
The frequency spectrum reveals several peaks in the amplitude corresponding to the frequencies of standing spin waves. 
For the Fourier transformation we integrate over a time interval of $100\,$ ps. Note that especially after $10\,$ ps lower frequencies are dominating. 
Therefore, the amplitudes of low frequencies are larger than those of higher frequencies. 
%Here, we consider a system without temperature and we obtain high frequency modes up to a few THz. In experiments, frequency up to around $1\, \mathrm{THz}$ has been observed,~\cite{Razdolski2017} but due to the temperature in the system low amplitudes at higher frequencies could not be resolved there. Furthermore, we use the signal from only the last layer, whereas in experimental measurements signal from deeper inside averages out parts of the magnons and therefore, the signal especially at higher frequencies is reduced.

Standing spin waves are formed if the thickness of the system is a multiple of the wavelength and the wavevector $q_z$ is given by: 
\begin{align}
    \label{eq_standing}
    q_z=\frac{\pi \cdot n}{d_2}\mathrm{,}
\end{align}
% !!! we need to reconsider some variables to avoid two different variables having the same name -P
% Ulrike: replaced d and l with d_2 (as defined in methods as thickness for second FM); More conflicts?
where $d_2$ is the thickness of the second ferromagnet as introduced before and $n=0,1,2,\dots$ denote the mode. 

%To identify the corresponding frequencies, we determine the dispersion relation of the magnons numerically by simulating a larger system with a thickness of $d_2 = 400\, a \approx 115\, \mathrm{nm}$ and use a larger damping constant of  $\alpha=0.05$. We excite monochromatic spin waves by a coherent precession of an external magnetic field which couples only to the first layer. This method allows to stimulate one-dimensional monochromatic spin wave propagation in $z$-direction. We perform simulations with monochromatic spin waves varying the excitation frequency and determine the wave vector by fitting the spatial profile to a plane wave with exponential decay with increasing propagation distance. The obtained results are shown in Figure~\ref{fig5}. The numerical data are shown as dots in the figure. 

%\textbf{KC:create another subsection here?  magnon frequency calculation}
%Comment UR: I think the parts would be too short and I would prefer to not create another section. Otherwise, we should restructure the parts on a larger scale.

To obtain the corresponding frequencies, we determine the dispersion relation of the magnons analytically. For that purpose we consider the linearized LLG equation without damping and solve it analytically as described in Ref.\ \cite{Ritzmann2014}.
The solution of the resulting coupled set of equations are plane waves and
we obtain the Kittel formula \cite{Herring1951}:
\begin{align}
\label{eq_kittel}
    \hbar\omega=\sqrt{(2k_x+J_\mathrm{eff}(\mathbf{q}))\cdot(2k_x-2k_z+J_{\mathrm{eff}}(\mathbf{q}))}\textrm{,}
\end{align}
where $J_\mathrm{eff}(\mathbf{q})$ denotes the total contribution of exchange interaction and is given by
\begin{align}
    J_\mathrm{eff}(\mathbf{q})=\sum_{k=1}^6{J_k\cdot\left(N_k-\sum_{\theta_k} 2\cdot\cos{(\mathbf{q}\cdot\boldsymbol\theta_k)} \right)}\textrm{,}
\end{align}
where $J_k$ is the exchange interaction with the $k$-th neighbor, $N_k$ is the number of $k$-th neighbors and $\theta_k$ are the distance vectors between the magnetic moment and the considered neighbor.
We consider only spin-wave propagation in $z$-direction with $\mathbf{q}=q_z\mathbf{\hat{z}}$ and simplify the expression by approximating $\cos{x} \approx 1 - (1/2)\,x^2-(1/24)\,x^4$. We can then obtain a simplified expression for the exchange interaction
\begin{align}
    \label{eq_ex}
    J_\mathrm{eff}&=J_\mathrm{eff,1}(aq_z)^2+J_\mathrm{eff,2}(aq_z)^4
\end{align}
with
\begin{align}
    J_\mathrm{eff,1}&= J_1+J_2+4J_3+11J_4+4J_5+4J_6,\\
     J_\mathrm{eff,2}&=\frac{1}{48}(J_1+4J_2+16J_3+83J_4+16J_5+64J_6).
\end{align}
\begin{figure}[t]
    \centering
    \includegraphics[width=0.49\textwidth]{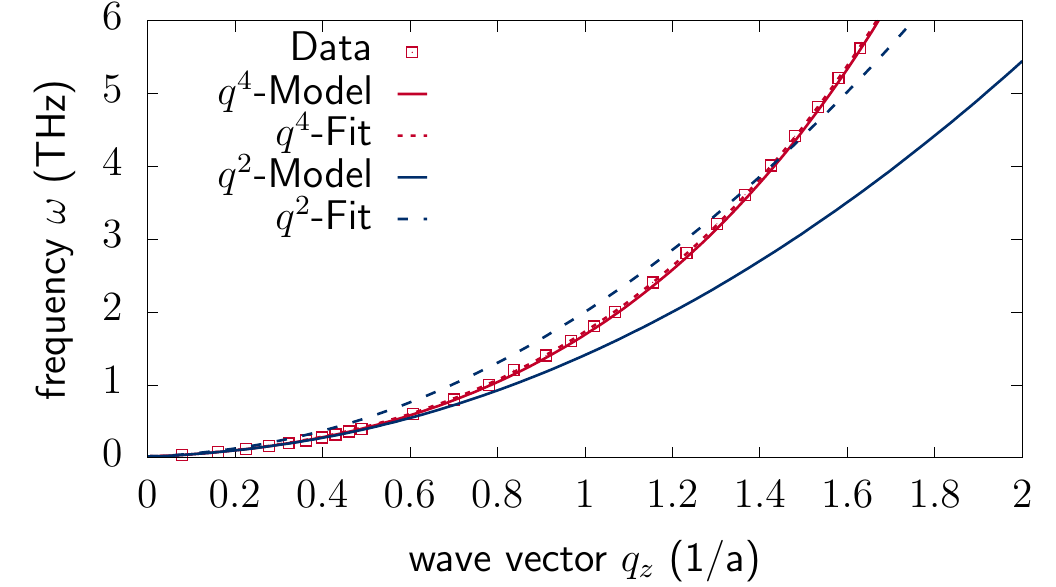}
    \caption{\label{fig5} Frequency of magnons as a function of the wavevector for one-dimensional spin-wave propagation. The red points show data obtained by our numerical simulations whereas the red and blue line show the analytical model based on equations (\ref{eq_kittel}) and (\ref{eq_ex}). The dotted red line shows a fit to the data points.}
\end{figure}
Making this approximation, we obtain for the effective exchange constants $J_\mathrm{eff,1}=5.60$\,meV and $J_\mathrm{eff,2}=1.42$\,meV. Note that especially the value $J_\mathrm{eff,1}$ is much smaller than the nearest neighbor exchange interaction. 
%Ulrike: Comparison with experiments does not work well. D=280meV*Angstrom^2?! But there is a unit problem!! 
In Figure \ref{fig5} we show the analytical model in comparison to numerical data obtained by simulating monochromatic spin waves. The analytical result including a $q^4$-term of the exchange shows a good agreement with the numerical data, whereas the approximation with only $q^2$ clearly deviates. A numerical fit of the data with a $q^2$-term gives a much higher effective exchange constant of $J_\mathrm{eff,1}^{\mathrm{fit}-q^2}=7.90$\,meV. The fitting curve can describe overall the numerical results, but clear deviations occur. The obtained fitting parameter strongly deviates from the analytical model, demonstrating the importance of the higher order corrections. A fit including the $q^4$-term shows small deviations at higher frequencies above $5\,$THz. In order to describe the dispersion relation with a high precision, we also fit the dispersion relation and obtain small corrections for $J_\mathrm{eff,1}$ and $J_\mathrm{eff,2}$.  The fitted values are $J_\mathrm{eff,1}^\mathrm{fit}=5.72$\,meV and $J_\mathrm{eff,2}^\mathrm{fit}=1.14$\,meV.

\begin{figure}[t]
    \centering
    \includegraphics[width=0.49\textwidth]{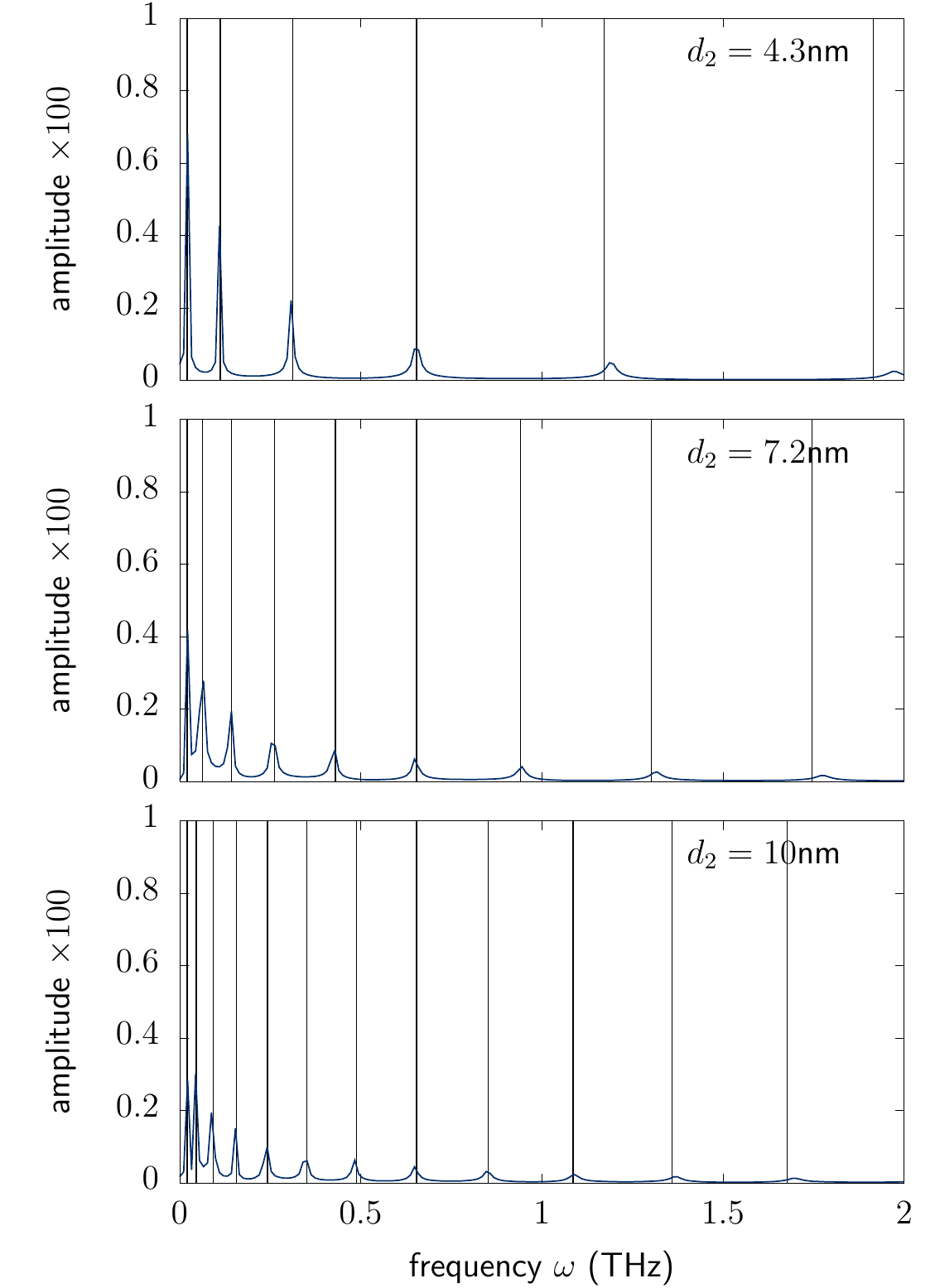}
    \caption{\label{fig6} Magnon frequency spectra for different thicknesses of FM2. The blue lines show the spin-wave amplitude as a function of the frequency obtained by Fourier transformation in time of the magnetization of the last layer. The perpendicular black lines illustrate the predicted peak position for standing waves given by equation (\ref{eq_standing}).
    }
\end{figure}
With this analytical model, we predict the spin-wave peak positions using the conditions for standing waves given by equation (\ref{eq_standing}). In Figure~\ref{fig6}, we show the frequency spectra of the excited spin waves up to frequencies of $2\,$THz for different thicknesses obtained by Fourier transformation as before. 
%In all cases, we use the averaged magnetization of the last layer and perform a Fourier transformation in time for a time interval of 100ps. 
For a thickness of $d= 4.3\, \mathrm{nm}$ 
%In the upper panel of Figure \ref{fig6}, a thickness of $d= 4.3\, \mathrm{nm}$ is shown within a frequency spectra of up to $2\, \mathrm{THz}$ and 
one sees a strong peak at $q=0$ (peak with the lowest energy) corresponding to the FMR mode and 4 further peaks. The FMR peak appears in all shown cases at the same frequency, but by increasing the thickness of the film, the number of peaks within the $2\,$THz range is increasing as shown for thicknesses of $7.2\,$nm and $10\,$nm. The predicted peak positions, which are illustrated as perpendicular lines, are in very good agreement with the numerically observed peaks. Note that the amplitudes of the single frequencies is decreasing for increasing thickness. The results demonstrate that in the time regime of $100\,$ps, only standing spin waves fulfilling equation (\ref{eq_standing}) are relevant.
% and the prediction are in very good agreement with numerical results. The dispersion relation of the involved magnons can be well described by linear spin wave theory and it strongly deviates from considering only nearest neighbor exchange due to strong frustration effects. 
Frequencies up to a few THz are excited through the femtosecond STT with a pulse duration of the excited hot electron spin current of about $500\, {\rm fs}$.

\subsection{Magnon distributions for different STT penetration depths}

\begin{figure}[t]
    \centering
    \includegraphics[width=0.49\textwidth]{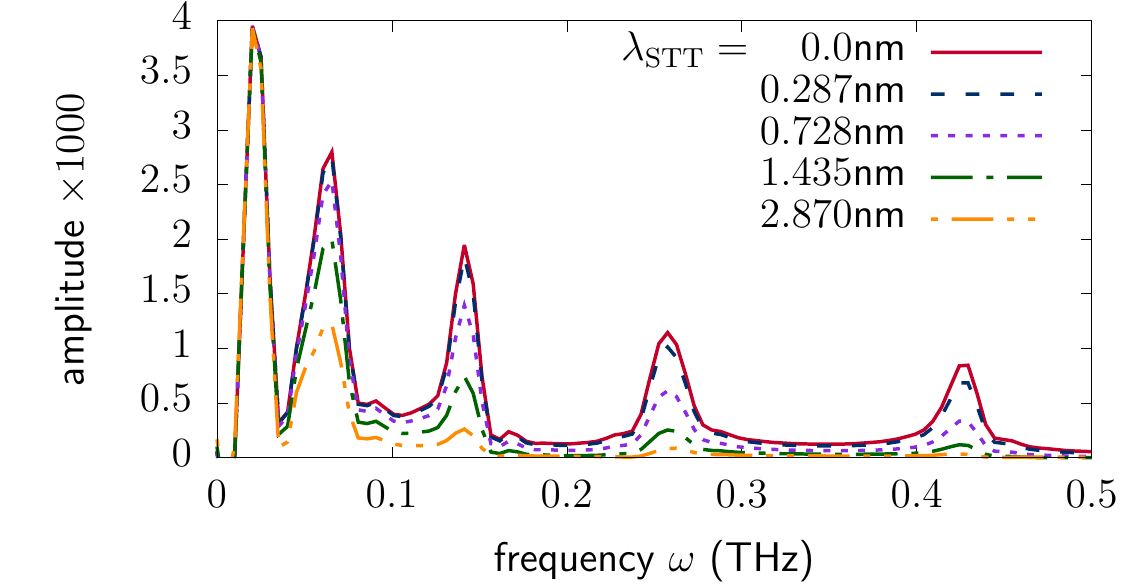}
    \caption{\label{fig7} Calculated magnon amplitudes as a function of the frequency for different penetration depths $\lambda_\mathrm{s}$.}
\end{figure}
To provide a more realistic description of the spin dynamics, we now consider a finite penetration depth of the transverse spin. Large penetration depths suppress high frequency magnons and therefore, recent experiments investigated the maximum frequencies that can be excited to determine the upper bound of the penetration depth. Here, we will explore the modifications of the frequency-dependent amplitudes of the excited magnons due to a finite penetration depth.
We perform simulations assuming an STT that enters the film with a characteristic penetration depth as described by equation (\ref{spatial_STT}).  As for the case of full absorption at the interface, we use a film thickness of $d_2 = 25\, a = 7.2\, \mathrm{nm}$ for the second ferromagnet and perform our numerical simulation for a time interval of 100\,ps for the average magnetization of the last layer.  We study the influence of the penetration depth in a range of $\lambda_\mathrm{s}$ equal to zero up to almost $3.0\, \mathrm{nm}$.  

The resulting spin-wave amplitudes as a function of the frequency for different penetration depths are shown in Figure~\ref{fig7}. For comparison, we also include the obtained data from zero penetration depth, $\lambda_\mathrm{s}=0$. The position of the peaks are the same in all cases, as shown in the figure, but the amplitudes for each peak strongly differ. The amplitude of the FMR mode, corresponding to $q_z=0$, is almost the same for the different penetration depth. On the other hand, already at the first spin-wave mode, the amplitude decreases significantly for the highest penetration depth. The amplitude is reduced by more than a factor of two for the largest value of $\lambda_\mathrm{s}$ and here the penetration correspond to 1/5 of the wavelength of the mode. For the largest penetration depth modes with  $n\geq4$  are no longer excited. The penetration depth in that case is larger than half of the wavelength of the magnon modes. For smaller values of the penetration depth, higher modes are still excited, but the amplitudes of the modes are strongly suppressed. A large penetration depth leads to an almost complete suppression of magnons with a wavelength $\lambda\leq 2\lambda_\mathrm{STT}$, but they can also cause a strong reduction of the amplitudes of the magnons with larger wavelength than $2\lambda_\mathrm{STT}$. Although these results are in general agreement with experimental observations, it shows that the determination of an upper limit for $\lambda_\mathrm{STT}$ is rather difficult, due to a strong suppression for all modes with finite wavelength.

\subsection{Time evolution of the frequency spectra}
\begin{figure}
    \centering
    \includegraphics[width=0.49\textwidth]{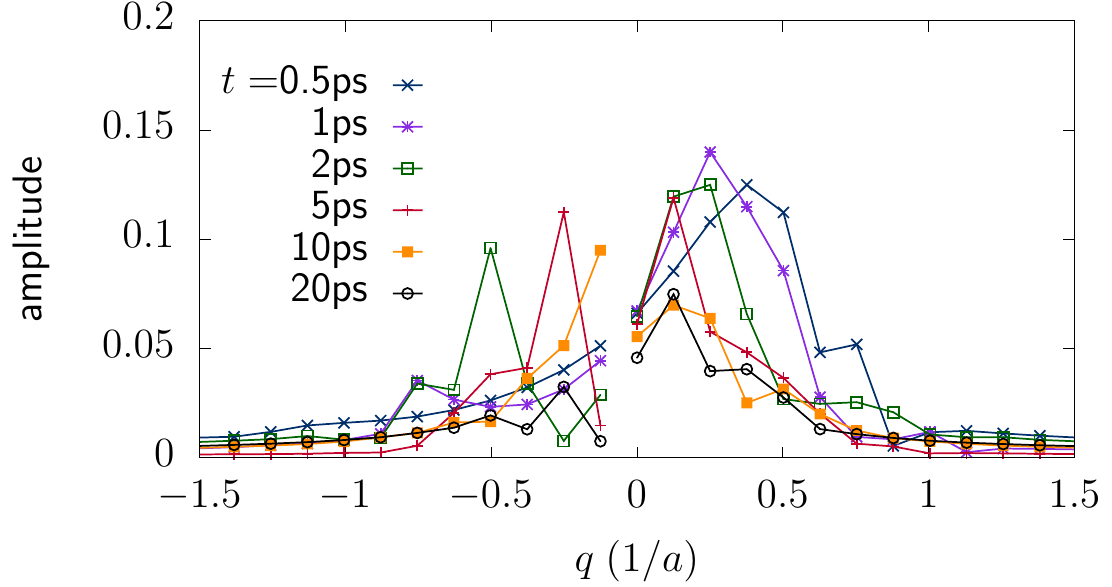}
    \caption{Calculated spin-wave amplitudes as a function of wavevector $q_z$ obtained by space-domain Fourier transformation at different times.}
    \label{fig8}
\end{figure}

As next step, we investigate the temporal evolution of the excited magnon spectra. Instead of calculating a Fourier transformation in the time-domain, we perform a Fourier transformation now in the space domain and obtain the amplitudes as a function of the wavevector of the modes. The results are shown in Figure \ref{fig8}. At 500\,fs, a broad spectrum of frequencies with mainly positive wave vectors is propagating through the system. At this times the spin waves did not reach the end of the system and therefore no reflection has occurred. The data shows that the maximum in the amplitudes occurs around $q=0.75(1/a)$, which corresponds to a frequency of about 1\,THz. But also higher modes are excited with wave vectors of $q=1.5(1/a)$ and a frequency of about 4.5\,THz. At 1\,ps, the maximum amplitude is already shifted to lower wavevectors and the highest frequencies are reflected and contributions with negative wave vectors are forming. At later times, beyond 5\,ps, one can clearly see that modes with lower wavevectors starts to dominate and spin waves with larger wavevectors, and consequently higher frequencies are decaying strongly within the first 10\,ps.

As last step, we want to compare our observation to the lifetimes of the frequencies due to Gilbert damping. To describe the frequency dependent lifetimes due to Gilbert damping in the magnetic layer, we consider again the linearized LLG equations, but now we include also the damping contributions. We solve the coupled set of equation considering plane waves as solution of the system. The imaginary part of the eigenvalue corresponds to the frequency of the system and the real value is linked to the frequency-dependent lifetime of the spin waves. Note that the damping term also modifies the frequency of the magnons, which however only becomes relevant for larger damping values. The corrected frequency $\omega'$ is then given by:
\begin{align}
    \hbar\omega'=\sqrt{(\hbar\omega)^2-\left(\frac{\alpha}{1+\alpha^2}\right)^2\left(J_\mathrm{eff}(\mathbf{q}+2k_x-k_z\right)}
\end{align}
The lifetime $\tau$ of the spin waves due to Gilbert damping describes an exponential decay of the modes after excitation and is strongly frequency dependent. We obtain for the wavevector dependent lifetime:
\begin{align}
    \label{eq_lifetime}
    \tau(\mathbf{q})=\frac{1}{\frac{\alpha}{1+\alpha^2}\left(J_\mathrm{eff}(\mathbf{q})+2k_x+k_z\right)}
\end{align}
\begin{figure}[t]
    \centering
    \includegraphics[width=0.49\textwidth]{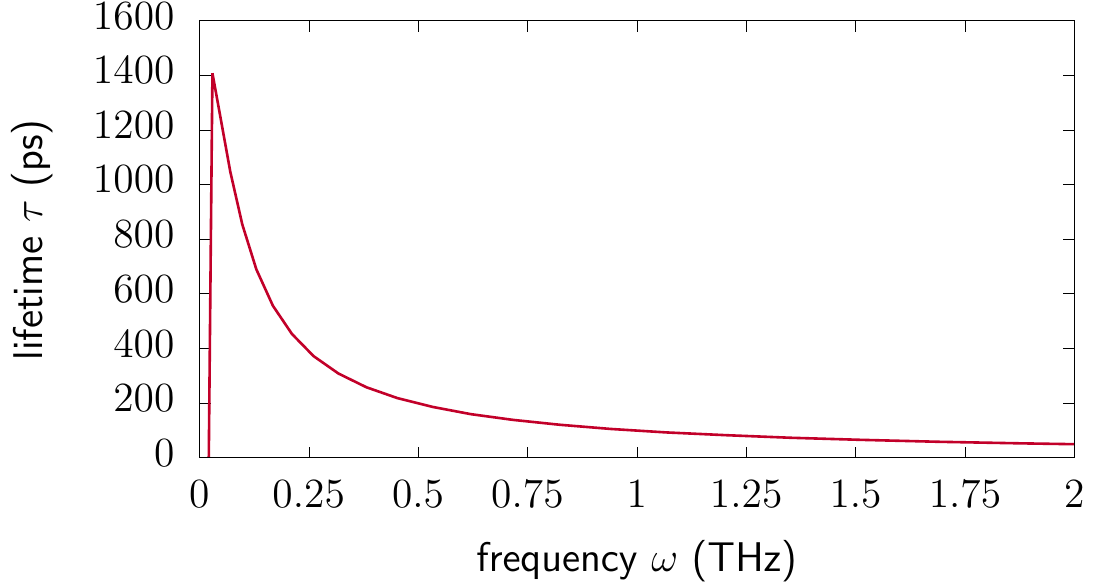}
    \caption{\label{fig9} Lifetime of magnons as a function of the frequency given by equation (\ref{eq_lifetime}) and a damping constant of $\alpha=0.01$.}
\end{figure}
The resulting lifetime as function of the frequency is shown in Figure \ref{fig9}. Since we use a very small damping value of $\alpha=0.01$, we neglect the frequency correction and use $(1+\alpha^2)\approx 1$. The lifetime for larger frequencies is significantly reduced compared to the lifetime of magnons with low frequencies. Magnons with frequencies above 1\,THz decay within 100\,ps, whereas magnons with lower frequencies can have lifetimes above 1\,ns.

The long lifetimes for the spin waves with low frequencies are in good agreement with our numerical observations and also with experimental findings. But although spin waves with higher frequencies have a much lower lifetime than the ones with lower frequencies, our numerical results indicate an even faster decay of these high-frequency modes. This indicates that nonlinear processes occur and magnon-magnon interaction leads to a stronger decay of the modes with higher frequencies.
%One possible reason for this observation could be magnon-magnon interaction especially appearing due to the multiple reflections at the boundary. This decay mechanism is not included in the lifetimes due to Gilbert damping.
%This is in agreement with our results shown in the Figures~\ref{fig3} and \ref{fig4}, in which we describe fast oscillations due to high frequencies in the first picoseconds and much slower oscillations and increasing wavelength above 10 picosecond. The large lifetime of the lowest frequencies are in agreement to our numerical observations and experiments, in which magnetization dynamics during $500\,{\rm ps}$ have been observed.

\section{Conclusion}
\label{Sec:Conclusions}
To summarize, we have studied theoretically the laser-induced excitation of THz magnons in Fe$|$Cu$|$Fe trilayer structures with two ferromagnetic materials having perpendicular magnetization orientations. We use a theoretical model to describe superdiffusive hot electron transport leading to ultrafast spin-current transfer. This spin-current bursts excite THz magnons in the adjacent second ferromagnet via an ultrafast spin-transfer torque. We combine our model with atomistic spin-dynamics simulations including exchange interactions beyond nearest neighbor interactions to describe the magnetization dynamics in the second ferromagnet on the atomic scale. In this way, we demonstrate the excitation of THz magnons and the formation of standing spin waves within the first picoseconds, as well as the larger lifetimes of the lowest frequency modes in good agreement with experimental observations. We analyze how the magnonic distribution depends on the penetration depth of the spin-transfer torque and we find a complete suppression of magnons with wavelength $\lambda\leq 2\lambda_\mathrm{STT}$. The decrease of magnon mode population with frequency (studied within the first 100 ps) is apparently significantly faster for longer penetration depths. We confirm that STT penetration depth has to 
be smaller than 1 nm in order to achieve a significant occupation of the \textbf{4th} magnon mode, as observed experimentally \cite{Razdolski2017}. 
%Comment UR: We cannot confirm anything, since we don't know how many modes are excited. My conclusion would be that the penetration depth is even shorter than the given limit in experiments, since all modes are suppressed and the appearance of higher modes requ
%\textbf{ We confirm that STT has to be confined within 1 nm from the interface.}

Our results demonstrate that laser-induced hot electron spin currents offer a new pathway to excite high-frequency magnons in the THz regime, which allows for new design concepts for ultrafast spintronics and high-frequency magnonics applications.  The developed theory can be used to tailor the trilayer composition so that the desired magnonic contribution is enhanced. 
		
\begin{acknowledgments}
We thank J.\ Hurst and A.\ Melnikov for valuable discussions.
This work was supported by the Deutsche Forschungsmeinschaft via grants RI 2891/1-1 and RI 2891/1-2, by the Swedish Research Council (VR), and by the European Regional Development Fund 
in the IT4Innovations national supercomputing center - path to exascale project 
(project number CZ.02.1.01/0.0/0.0/16\_013/0001791) within the Operational Programme Research, Development and Education
and by the Czech Science Foundation for support (grant number 18-07172S). We further acknowledge support from the K.\ and A.\ Wallenberg Foundation (grant No.\ 2015.0060) and the Swedish National Infrastructure for Computing (SNIC).
This work was supported by The Ministry of Education, Youth and Sports from the Large Infrastructures for Research, Experimental Development and Innovations project "IT4Innovations National Supercomputing Center – LM2015070".
\end{acknowledgments}
% I just copied the acknowledgement from the domain wall manuscript. Is it OK? -- Pavel

\appendix

\section{Exchange interactions}
\label{App:exch_int}
Exchange interaction constants can be calculated using infinitesimal spin rotations \cite{Liechtenstein1987} or frozen spin waves total energy calculations \cite{Halilov1998}. 
Here we use the expression of Liechtenstein \textit {et al.} which reads
\begin{equation}
  \begin{split}
    J_{ij} = & \frac{1}{\pi}\,\Im \int_{-\infty}^{E_{F}}{\rm d}E\int_{\Omega_{i}}{\rm d}\br\int_{\Omega_{j}}{\rm d}\br' \\
             & \Bxc\left(\br\right) \Gup\left(\br, \br', E^{+} \right)\, \Bxc\left(\br'\right) \Gdn\left(\br', \br,E^{-}\right)\,,
  \end{split}
\label{Eq:Liecht}
\end{equation}
\begin{figure}[t!]
    \includegraphics[width=.98\columnwidth]{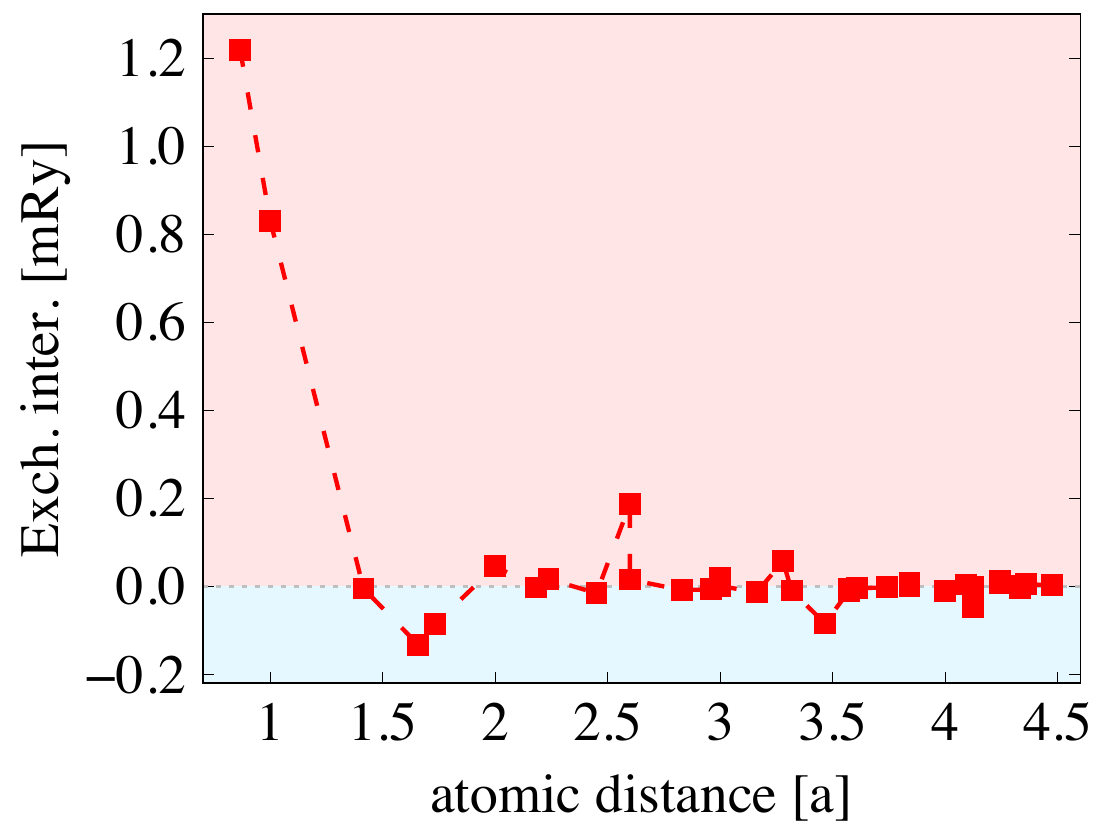}
    \caption{Exchange interactions for the 6 nearest neighbours calculated for bcc Fe lattice as a function of the distance between the atoms. The distance is given in the units of the Fe lattice constant, $a$.}
\label{Fig:exch_int}
\end{figure}
where $G^{\sigma}$ is the spin-dependent Green function with spin $\sigma \in \{ \uparrow, \downarrow\}$, 
$\Bxc$ is the magnetic field from exchange-correlation potential, $\Omega_i$ is volume of sphere with center in $i$-th atom position,
and $E^{\pm}= \lim_{\alpha\rightarrow0}E \pm i\,\alpha$ with $i = \sqrt{-1}$. Fig.~\ref{Fig:exch_int} shows the obtained exchange interactions for a bcc Fe lattice.

In our atomistic spin dynamics simulations, we include exchange interaction up to the $6{\rm th}$ neighboring shell, which corresponds to a distance of the neighbors of up to $2a$.
%%%

\bibliography{./library}

\end{document}